\documentclass[aps,prl,twocolumn,amsmath,amssymb,floatfix,showpacs]{revtex4}
\usepackage{epsfig,psfrag,subfigure}
\usepackage{graphicx,psfrag,subfigure}
\usepackage{color}
\usepackage{bm}

\newcommand\beq{\begin{equation}}
\newcommand\eeq{\end{equation}}
\newcommand\bea{\begin{eqnarray}}
\newcommand\eea{\end{eqnarray}}
\newcommand\non{\nonumber}

\newcommand\ig{\includegraphics}
\newcommand\bib{\bibitem}

\begin{document}

\title{Correlations and beam splitters for quantum Hall anyons}
\author{Smitha Vishveshwara$^1$ and N. R. Cooper$^2$}
\affiliation{$^1$ Department of Physics, University of Illinois at Urbana-Champaign,
1110 W. Green St., Urbana, IL 61801-3080, U.S.A.\\
$^2$T.C.M. Group, Cavendish Laboratory, J. J. Thomson Ave., Cambridge
CB3 0HE, U.K.}
\date{\today}

\begin{abstract}
We study the system of two localized anyons in the lowest Landau level
and show how anyonic signatures extrapolating between anti-bunching
tendencies of fermions and bunching tendencies of bosons become
manifest in the two-particle correlations. Towards probing these correlations,
 we discuss the influence of a saddle potential on
these anyons; we exploit analogies from quantum optics to analyze the
time-evolution of such a system. We show that the saddle potential can
act as a beam splitter akin to those in bosonic and fermionic systems
and can provide a means of measuring the derived anyonic signatures.
\end{abstract}
\maketitle

 The quantum statistics of bosons and fermions plays a fundamental role
over a vast range of length scales and diverse physics such as the spatial
distribution and energetics of electrons in atoms, certain constraints on
scattering cross sections in nuclear physics, the existence of superfluids and
the stability of neutron stars. Statistical signatures key to these phenomena
have been explored over the past decades via analyses of two-particle
correlations including in seminal studies of bosonic bunching properties by
Hanbury Brown and Twiss \cite{HBT} and Hong {\it et al.}\cite{Hong}. The latter, which
also has its fermionic counterpart \cite{Loudon, Schonen,Yamamoto}, performs
time-resolved coincidence measurements of pairs of photons incident on a beam
splitter from two uncorrelated sources collected at two detectors. While these
analyses have established the allowed quantum nature of particles in three dimensions,
the past decade has drawn attention to the study of two-dimensional ``anyons'',
quasiparticles which obey fractional statistics interpolating between those of fermions
and bosons \cite{leinaas,wilczek}. Given the current rapid experimental
progress in two-dimensional systems and the keen quest for topologically
ordered states which can be ascertained by the detection of anyons, a
fundamental understanding of these entities analogous to that of fermions and
bosons is much called for.

  Here, we explore the (anti)bunching properties of a system of two non-interacting
 Abelian anyons by analyzing the behavior of specific observables and propose a means of realizing a beam splitter wherein these anyonic
 properties become manifest. The common wave function for these anyons by definition
 picks up a phase of $e^{i\pi\alpha}$ ($e^{-i\pi\alpha}$) upon a
anticlockwise (clockwise) exchange of the particles\cite{leinaas}. The
parameter $\alpha$ lies in the range $0 < \alpha \le 1$; $\alpha=0$ and $1$
correspond to bosons and fermions respectively. Of direct relevance to bulk
quasihole excitations
 in the quantum Hall system\cite{halperin,laughlin1,leinaas,kjonsberg_leinaas} --
 a paradigm for anyonic statistics --
 we study a two-dimensional system of two anyons
  in a magnetic field projected onto the lowest Landau level(LLL).(In particular,
 for Laughlin states \cite{laughlin1}, quasiholes have fractional charge $q=-e/m$
 and statistics $\alpha=1/m$, where $m$ is an odd integer
 \cite{laughlin2,arovas}.)
 We find that while anyonic signatures in the LLL are subtle, they clearly extrapolate
between their bosonic and fermionic counterparts. We show that the presence of
a saddle potential offers a means for LLL anyons to approach one another along
two incoming limbs and then propagate away along two out-going
 limbs, akin to the photonic beam splitter settings, and that analogous coincidence
 measurements made along the limbs can reflect our predicted anyonic signatures.

The Hamiltonian for two anyons in a perpendicular magnetic field ${\bm B}=
B\hat{{\mathbf k}}$ has the decoupled form 
\bea \hat{H} &=& \frac{1}{4\mu}\left(\hat{P}_x +
\frac{qB}{c} \hat{Y}_c\right)^2 +
\frac{1}{4\mu} \left(\hat{P}_y - \frac{qB}{c} \hat{X}_c\right)^2 \non \\
& & + \frac{1}{\mu} \left(\hat{p}_x + \frac{qB}{4c} \hat{y}_r\right)^2 + \frac{1}{\mu}
\left(\hat{p}_y - \frac{qB}{4c} \hat{x}_r\right)^2 , \label{eq:hamQH} \eea in terms of
center of mass (c.o.m.)  and relative variables.
Here the anyons are assumed to have mass $\mu$ (which is immaterial
when states are projected to the LLL) and charge $q$. We have chosen
the symmetric gauge 
${\bm A}_\gamma=(B/2) (-y_\gamma\hat{\mathbf i}+x_\gamma\hat{\mathbf j}),\gamma=1,2$ for each particle. The
c.o.m. coordinate and momentum are given by
${\bm R}_c=({\bm r}_1+{\bm r}_2)/2$ and ${\bm P}={\bm p}_1+{\bm p}_2$,
while the relative coordinate and momentum are given by
${\bm r}={\bm r}_1-{\bm r}_2$ and
${\bm p}=({\bm p}_1-{\bm p}_2)/2$. In both the c.o.m. and relative
coordinate sectors, the LLL is spanned by energy degenerate angular
momentum eigenstates.  The c.o.m Hilbert space is identical to that of
a single particle; angular momentum states $|n\rangle_c$ are
eigenstates of angular momentum $\hbar A^{\dagger}A$ having
eigenvalues $n\hbar$,where $n$ is an integer \cite{jain}.  Here, the
usual commutation rules $[A,A^{\dagger}]=1$ are satisfied and the
components of the guiding centers have the form
$\hat{X}=l(\hat{A}+\hat{A}^{\dagger})/2$ and
$\hat{Y}=il(\hat{A}-\hat{A}^{\dagger})/2$,
where $l=\sqrt{\hbar c/qB}$
is the single-particle magnetic length\footnote{Note that the natural
magnetic lengths in the centre-of-mass and guiding center co-ordinates
are $l_c = l/\sqrt{2}$ and $l_r= \sqrt{2}l$, where $l$ is the single
particle magnetic length.}
In the relative coordinate sector, the anyon boundary condition is not respected by the guiding center coordinates, $\hat{x},\hat{y}$, but it is by
their quadratic combinations,
$\hat{a}\equiv (\hat{x}^2+\hat{y}^2)/8l^2$,
$\hat{b}\equiv (\hat{x}^2-\hat{y}^2)/8l^2$,
$\hat{c}\equiv (\hat{x}\hat{y}+\hat{y}\hat{x})/8l^2$.
These operators respect a $sp(1,R)$ algebra\cite{hansson}. 
The relative coordinate Hilbert space consists of irreducible
representations of this algebra $|k,\alpha\rangle_r$, where $k$ is an integer, and
correspond to eigenstates of the angular momentum 
$\hat{L}=\hbar (2\hat{a}-1/2)$ having
eigenvalues $(2k+\alpha)\hbar$~\cite{hansson,kjonsberg_leinaas}. 

 Localized anyons can be composed as linear combinations of the degenerate LLL
 angular momentum states; Ref.~\cite{kjonsberg_leinaas} provides a careful derivation of
 the form of such anyon-coordinate states that most closely describe  localized LLL quasiholes, leading to the expected exchange statistics. These states can be decomposed into product states of localized
 states centered at the dimensionless c.o.m. coordinates $Z=(z_1+z_2)/2$ and relative coordinates 
 $z=z_1-z_2$, where the individual anyons are centered at
 $z_\gamma=(x_\gamma+iy_\gamma)/l, \gamma=1,2$. These localized states have
 the form\cite{kjonsberg_leinaas}\footnote{For the relative coordinates, these states are not equivalent to coherent states except for the fermionic and bosonic cases 
 but they do exhibit the same asymptotic behavior.}
\bea |Z\rangle_c &=& e^{-|Z|^2/2}\sum_{n=0}^{\infty}\frac{(Z^*)^n}{\sqrt{n!}}|n\rangle_c , 
 \label{eq:localizedZ}
\\
|z\rangle_{\alpha} &=&
N_{\alpha,z}\sum_{k=0}^{\infty}\frac{(z^*/2)^{2k+\alpha}}{\sqrt{\Gamma
(2k+\alpha+1)}}|k,\alpha\rangle_r. \label{eq:localized} \eea
Here, we express
the normalization $N_{\alpha,z}$, which in itself contains information on
statistics, in terms of a sum of two confluent hypergeometric functions as 
$2|N_{\alpha,z}|^{-2}\Gamma(1+\alpha)=(|z|/2)^{2\alpha}[M(1,1+\alpha,|z|^2/4)+M(1,1+\alpha,-|z|^2/4)]$.
 The convention chosen for the relative co-ordinate localized state explicitly respects the
anyonic boundary condition in picking up the desired phase under the exchange
action $z\rightarrow ze^{i\pi}$.

The localized states defined above are consistent with the expected forms for
fermions and bosons. For these cases, (anti)symmetrization is brought about by
the construction
\begin{equation}
|z\rangle_{1/0}=e^{|z|^2/8}N_{1/0,z}[|z\rangle_d\mp|-z\rangle_d],
\label{eq:fblocalized}
\end{equation}
where $|z\rangle_d$ refers to the localized state form for distinguishable
particles, analogous to Eq.\ref{eq:localizedZ} but with $Z \to z/2$ in view of the change in effective magnetic length[23]. In fact, for  localized states, the probability density must be symmetrically peaked close to both relative coordinates $z$ and $-z$ for all indistinguishable particles; this can be ascertained by analyzing the anyon state of Eq. \ref{eq:localized}. 
The construction (\ref{eq:fblocalized}) explicitly shows that the fermion/boson boundary
conditions allow only odd/even angular momentum states in the relative
co-ordinate localized state decomposition.   By evaluating the overlap between
$|z\rangle_d$ and $|-z\rangle_d$, it can be shown that the normalization
constant has the simple limiting forms $|N_{1,z}|^{-2}=\sinh(|z|^2/4)$ and
$|N_{0,z}|^{-2}=\cosh(|z|^2/4)$, respectively.

 One of the most direct measures of observing quantum statistics and related
 (anti)bunching behavior is the average guiding center separation squared,
 $\langle \hat{{\mathbf r}}^2\rangle \equiv 
 \langle \hat{x}^2 + \hat{y}^2\rangle$. 
It is well known that for any
 generic system of spinless fermions/bosons, (anti)symmetrization leads to this
 average separation being measurably greater/smaller than the value for distinguishable
 particles\cite{baym, griffiths}. Generally, this statistical effect becomes most pronounced at
 smaller separation while at larger separation, statistical correlations decay
 out in a manner characteristic to the particular system and the average
 separation for distinguishable and indistinguishable particles coincide. Here, to
 quantify this statistical effect, we define a bunching
  parameter
  \begin{equation}\chi(|z|,\alpha)\equiv \frac{1}{4l^2}\left[_{\alpha}\langle z|  \hat{{\bm r}}^2 | z \rangle_{\alpha}-\; _{d}\langle z|  \hat{{\bm r}}^2|z\rangle_d\right],
  \label{eq:bp}\end{equation}
where the factor of $4l^2$ is a matter of convention. A
(positive)negative value of $\chi$ implies (anti)bunching in
comparison with distinguishable particles. 

 We now evaluate the bunching parameter for localized
  LLL anyons.
 For
distinguishable particles, we have the expected form 
$_d\langle z| \hat{{\bm r}}^2|z\rangle_d=(|z|^2+2)l^2$, where the non-zero minimum value reflects the
finite width associated with the minimum uncertainty in guiding  center
positions $x$ and $y$, characteristic of states in the LLL. .
For fermions and bosons,  $\langle  \hat{{\bm r}}^2\rangle$ can be directly evaluated
 using the definition in Eq.\ref{eq:fblocalized}.  The bunching parameter takes the forms
 $\chi(z,1)= (|z|^2/4)[\coth(|z|^2/4)-1]$ for fermions and $\chi(z,0)= (|z|^2/4)[\tanh(|z|^2/4)-1]$ for bosons.  In keeping with expectations, the bunching parameter is always
 positive/negative for fermions/bosons and decays exponentially towards zero
 for large $|z|$.

\begin{figure}[t]
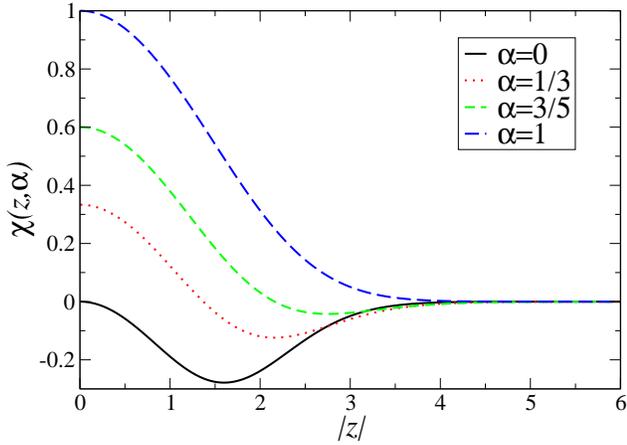
 \ig[width=3.25in]{chi_rescale.eps}
  \caption[]{(Color online)The bunching parameter as a function of the
    dimensionless distance between particles $|z|$ for different values of
    anyonic phase $\alpha$. Curves from the top-most along the $y$-axis
    correspond to values $\alpha=1$, $\alpha=3/5$, $\alpha=1/3$ and
    $\alpha=0$.} \vspace{-0.5cm}
\label{fig:bunchprop}
\end{figure}

 For anyons, the
 desired expectation values can be evaluated by using $ \hat{{\bm r}}^2 = 8l^2\hat{a}$, and the eigenstate property $\hat{a}|k,\alpha\rangle_r
 =(k+\alpha/2+1/4)|k,\alpha\rangle_r$  and hence
 \begin{eqnarray}
 _\alpha\langle z|8\hat{a}-2|z\rangle_{\alpha}&=&|N_{\alpha,z}|^2\sum_{k=0}^{\infty}
 \frac{(8k+4{\alpha})\left(\frac{|z|^2}{4}\right)^{2k+\alpha}}{\Gamma(2k+\alpha+1)} \nonumber \\
 & = &
   \frac{4\alpha \left[M(1,\alpha,\frac{|z|^2}{4})+M(1,\alpha,-\frac{|z|^2}{4})\right]}{M(1,1+\alpha,\frac{|z|^2}{4})+M(1,1+\alpha,-\frac{|z|^2}{4})}. 
 \label{eq:aexpval}
 \end{eqnarray}
Fig.\ref{fig:bunchprop} shows the trend exhibited by the bunching parameter
obtained from Eq.\ref{eq:aexpval}. Quite remarkably, the value of $\chi$ at
$|z|^2=0$ 
(which is not physically accessible in quantum Hall samples) 
directly 
reflects the statistical parameter; $\chi(0,\alpha)=\alpha$.
The limiting case of the fermion, as a function of $|z|$, $\chi(|z|,1)$
begins at a value of unity and then decays to zero in a monotonic fashion. For
the bosonic case, $\chi(|z|,0)$ always remains negative, beginning at zero
decreasing to a minimum value and then rising to taper towards zero. The
intermediate anyonic values of $\alpha$ interpolate between these two limiting
behaviors. For all anyons, $\chi(|z|,\alpha)$ begins at $\alpha$, decreases
below zero, reaches a minimum and finally tapers towards the zero. Hence, the
bunching parameter shows that all anyons exhibit anti-bunching at short length
scales and bunching at long scales and that the trend evolves continuously as a
function of $\alpha$. This result is surprising in
that one might expect $\alpha<1/2$ to be boson-like and $\alpha>1/2$ to be fermion-like. The behavior of
$\chi(|z|,\alpha)$ shown in Fig.\ref{fig:bunchprop} and its connection to
fractional statistics forms the heart of our results.

While the correlations discussed above bear distinct signatures of
   statistics, they are static in nature due to the projection to
   the LLL. In practice, probing these correlations requires
   endowing them with dynamics via the application of appropriate potentials that lift the LLL degeneracy. Here we
   propose the application
   of a saddle potential whose effect on each particle can be
   described by $\hat{H}_s=\sum_{\gamma=1,2}U\hat{x}_\gamma \hat{y}_\gamma$, $U>0$, where $Ul^2$ is much smaller than the Landau level spacing thus retaining the 
   LLL projection. 
In terms of LLL eigenstate solutions for a
    single particle\cite{Fertig}, it has been shown that the saddle potential acts as a beam splitter in that particles approaching the origin along the $x$-axis tend to scatter
   either along the positive or the negative $y$-axis. Moreover, for two particles, the potential has the
   advantage of  being separable in terms of the relative and center of mass
   motion\cite{Alex}, hence preserving the decoupling of these degrees of freedom, and of respecting the anyon boundary conditions.
   The saddle potential, when projected to the LLL\cite{jain,kjonsberg_leinaas}, can be expressed as
    \begin{equation}
    \hat{H}_s^P=\frac{1}{2}iUl^2[\hat{A}^2-(\hat{A}^{\dagger})^2] + 2 U l^2 \hat{c}
    \label{eq:HamSaddle}
    \end{equation}
 It can be shown that, as for the single particle case, eigenstates of this Hamiltonian correspond to scattering states in 
the relative and center of mass sectors and that anyonic statistics translates to scattering phase shifts in the relative 
sector\cite{Alex}. Here we show that for pairs of localized particles
   traveling along opposite limbs of the saddle on the $x$-axis, the choice of
   propagation along the $y$-axis directly reflects the correlations similar to those shown in
   Fig.\ref{fig:bunchprop}.

To gain an insight on the propagation of localized states along saddle
potentials, we present an analysis of single particle physics
exploiting analogies in quantum optics\cite{qtmoptics} which can also
be applied for the c.o.m. behavior. The associated localized
c.o.m. state in Eq.\ref{eq:localized} has the coherent state form
$|Z\rangle_c=\exp(Z\hat{A}^{\dagger}-Z^*\hat{A})|0\rangle_c\equiv
\hat{D}(Z)|0\rangle_c$. In the Schr{\" o}dinger picture, we can consider the
time-evolution of the coherent state due to the saddle potential:
$|Z(t)\rangle_c =e^{-i\hat{H}_s^Pt/\hbar}|Z\rangle_c$.  In the language of
quantum optics, the time-evolution operator has the form of the
squeeze operator $\hat{S}(\xi)=\exp[\xi(\hat{A}^{\dagger})^2/2-\xi^*\hat{A}^2/2]$, where
for the c.o.m. sector we have 
$\xi=-Utl^2/\hbar$. The squeeze parameter $\xi\equiv r e^{i\phi}$
corresponds to squeezing along the direction $\phi$ with associated
aspect ratio $r$; here the squeeze is of magnitude  $Utl^2/\hbar$ along
the $x$-axis. We can now invoke the identity $\hat{S}(\xi)\hat{D}(Z)=\hat{D}(Z\cosh
r+Z^*e^{i\phi}\sinh r)\hat{S}(\xi)$ and the fact that
$\hat{D}(\beta)\hat{S}(\xi)|0\rangle$ represents a squeezed state having squeeze
parameter $\xi$ centered at $\beta$\cite{qtmoptics}. Hence, the
time-evolved coherent state flattens along the $y$-axis and its center
follows the trajectory 
$(X e^{-Utl^2/\hbar},Ye^{Utl^2/\hbar})$,
where $(X,Y)$ is the initial position of the coherent
state. Consistent with semi-classical dynamics along equipotentials of
a saddle potential, the center obeys $X(t)Y(t)$ being a constant.
Furthermore, any state having the initial condition $Y(t=0)>0$ evolves
asymptotically towards $Y(t\rightarrow\infty)\rightarrow+\infty$ and
likewise for the lower quadrant.

For the relative motion of anyons the analysis presented above cannot
be directly applied as the associated $sp(1,R)$ algebra is rather
involved. However, we surmize a few common features and
explicitly derive time-evolved expectation values of relevant
observables. Given that the initial state probability
density for the relative coordinate is peaked at $z$ and $-z$, 
over time, we expect it to asymptotically be distributed in the
upper and lower quadrants in a manner which depends on the statistics
of the particles.
The functioning of the saddle potential as a beam splitter is best
 seen when two localized state anyons are placed along or close to the
 $x$-axis, diametrically across one another with respect to the saddle
 point origin. As a
 function of time, the particles approach one another and then get
 deflected along the $y$-axis. Whether or not they travel in the same
 direction (along either the positive or negative $y$-direction) or in
 opposite directions depends on the magnitude of $\langle
 \hat{Y}^2\rangle$ compared to that of $\langle \hat{y}^2\rangle$. In
 fact, the quantity analogous to those measured in photonic and
 electronic beam splitters is $\langle \hat{y}_1\hat{y}_2\rangle=\langle
 \hat{Y}^2-\hat{y}^2/4\rangle$; a positive/negative value of
 $\langle \hat{y}_1\hat{y}_2\rangle$ indicates that the anyons traveled out along
 the same/opposite limbs, thus exhibiting bunching/anti-bunching
 behavior. These correlations are analogous to those between reflected
 and transmitted currents in electronic beam
 splitters\cite{Schonen,Yamamoto}.

The desired time-evolved expectation values are most easily evaluated in the Heisenberg
representation. 
From the commutation relations of
$\hat{a},\hat{b},\hat{c}$\cite{hansson}, one finds the Heisenberg
equations of motion $d\hat{a}/dt = -2U l^2 \hat{b}$, $d\hat{b}/dt =
-2U l^2 \hat{a}$.  The solutions of these equations yield $
\hat{x}^2(t) =e^{-2Utl^2/\hbar} \hat{x}^2(0)$ and $ \hat{y}^2(t)
=e^{2Utl^2/\hbar} \hat{y}^2(0)$ for the relative
co-ordinates. Similarly, and consistent with the above discussion, one
finds $ \hat{X}^2(t) =e^{-2Utl^2/\hbar} \hat{X}^2(0)$ and
$\hat{Y}^2(t) =e^{2Utl^2/\hbar} \hat{Y}^2(0)$ for the
c.o.m. coordinates.  By straightforward but lengthy evaluations of the
expectation values of these operators on the initial state described by
$Z$ and $z$, we find that the correlator for measuring the relative
motion of particles along the $y$-axis has the form
\begin{equation}
\langle \hat{y}_1\hat{y}_2\rangle  =  l^2e^{2Utl^2/\hbar}\left[\mbox{Im}[Z]^2-\frac{1}{4}\mbox{Im}[z]^2-\frac{1}{2}\chi
+\delta
\right]
\label{eq:x1x2}
\end{equation}
where $\chi(|z|,\alpha)$ is the bunching parameter introduced in
Eq.\ref{eq:bp}.
The
function $\delta(z,\alpha)$ is a small correction, which takes a maximum value
of 0.018, vanishes for $z=0$ and is present due to the deviation of localized states from coherent states.
Hence, for anyons placed on the $x$-axis, the sign of $\langle \hat{y}_1\hat{y}_2\rangle$, or equivalently, whether the particles went into the same limb
or opposite limbs, is entirely determined by the statistics and the bunching
parameter, which in turn depends on initial conditions. Given the exponential
dependence of  $\langle \hat{y}_1\hat{y}_2\rangle$  on time, the saddle potential acts as a
beam splitter whose read out amplifies initial correlations. As discussed
above and shown in Fig.\ref{fig:bunchprop}, for bosons/fermions, $\chi$ is
always negative/positive, and we recover the well known result that
statistically correlated bosons/fermions are expected to travel along the
same/different limbs. For anyons, the sign of the bunching parameter directly
determines whether pairs of particles go into the same limb or opposite
limbs. Thus, for particles initially placed close together, particles
propagate into opposite limbs as do fermions while those placed further out
propagate into the same limbs; the transition point between these two
dramatically different possibilities depends on the fractional statistics
parameter. A clear-cut signature of anyons is that, unlike for fermions and
bosons, both possibilities are present and accessible by tuning initial
conditions.

In an actual measurement, one can envision initializing two
 quasiparticles in the quantum Hall bulk in two locally created
 potential minima, as has been observed for single
 quasiparticles\cite{Yacoby}, applying an appropriately oriented
 saddle potential and collecting quasiparticles along receiver limbs
 by way of other local potential traps. Averages over a range of
 bunching parameter values are expected for a stream of particle pairs
 and associated uncertainties in initial positions. Average
 separations can be varied by changing the field, and thus the
 magnetic length, within a quantum Hall state. Alternatively, one can
 study correlations between dilute beams of quasiparticles conveyed
 along quantum Hall edge states and brought together via pinching
 \cite{Moty,Parsa}; correlations described here would require the
 application of a saddle potential at the pinched region.  While a
 complete analysis of such a system would require connecting edge
 physics to the bulk, we expect some signatures predicted here to be
 robust.

In
 conclusion, we have presented fundamental correlations characterizing LLL anyons and distinguishing them from their fermionic and bosonic counterparts. We have proposed the application of a saddle potential as a means of realizing a quantum Hall beam splitter that can display these correlations and associated direct signatures of fractional
 statistics.

We would like to acknowledge P. Kwiat and D. Sen for illuminating discussions. This
work was supported by the NSF under the grant DMR 06-44022 CAR, by the CAS fellowship at UIUC, and by EPSRC Grant EP/F032773/1.

\end{document}